\documentclass[sigconf,nonacm]{acmart}

\AtBeginDocument{%
  }

\setcopyright{acmlicensed}
\copyrightyear{2026}
\acmYear{2026}
\acmDOI{XXXXXXX.XXXXXXX}

\acmConference[UIST '26]{ACM Symposium on User Interface Software and Technology}{date}{location}

\usepackage{csquotes}
\usepackage{multirow}
\usepackage{array}
\usepackage{subcaption}

\begin{document}

\title{Breaking the Curse of Knowledge: Designing Personalized Jargon Support for Real-Time Online Meetings}


\author{Yifan Song}
\affiliation{%
  \institution{University of Illinois Urbana-Champaign}
  \city{Urbana}
  \state{IL}
  \country{USA}}
\email{yifan33@illinois.edu}

\author{Yijun Liu}
\affiliation{%
  \institution{University of Illinois Urbana-Champaign}
  \city{Urbana}
  \state{IL}
  \country{USA}}
\email{yijun6@illinois.edu}

\author{Wing Yee Au}
\affiliation{
  \institution{Fujitsu Research of America}
  \city{Santa Clara}
  \state{CA}
  \country{USA}}
\email{wau@fujitsu.com}

\author{Hon Yung Wong}
\affiliation{
  \institution{Fujitsu Research of America}
  \city{Santa Clara}
  \state{CA}
  \country{USA}}
\email{awong@fujitsu.com}

\author{Brian P. Bailey}
\affiliation{%
  \institution{University of Illinois Urbana-Champaign}
  \city{Urbana}
  \state{IL}
  \country{USA}}
\email{bpbailey@illinois.edu}

\author{Tal August}
\affiliation{%
  \institution{University of Illinois Urbana-Champaign}
  \city{Urbana}
  \state{IL}
  \country{USA}}
\email{taugust@illinois.edu}


\begin{abstract}
Cross-disciplinary communication is often hindered by specialized language (i.e., jargon) and uneven background knowledge. Recent advances in speech-to-text and large language models make it possible to provide jargon support during online meetings, but generic support (i.e., defining the same terms for everyone) can overwhelm listeners with definitions they do not need. We present ParseJargon, a system for personalized jargon support in real-time online meetings. We begin with an initial prototype to probe the use of single-sentence user profiles for personalization. We conducted a controlled study and showed that even this minimal personalization enhanced listeners' comprehension and engagement over generic support because of more precise jargon identification. Guided by insights from participants' feedback, we refined the system with more advanced personalization techniques, including in-session user feedback and portable glossary-based profiles. We evaluated how these techniques can further improve jargon identification precision using data collected in the controlled study to simulate personalization over time. We also conducted a latency test, complemented by a lightweight deployment, to analyze the system's real-time capability and usability.
\end{abstract}

\maketitle

\section{Introduction}

In workplaces, professionals frequently face challenges conveying specialized knowledge to colleagues from other disciplines~\cite{duolingo_linkedin_report, axios_report, keelawat_nbguru_2023, weirup_what_2024}. For instance, a machine learning engineer might struggle to communicate the concept of ``embeddings'' to a compliance officer concerned with data privacy, while healthcare professionals might face challenges describing ``quasi-experimental designs'' to policymakers without medical expertise. Such gaps in communication caused by domain-specific jargon limit interdisciplinary innovation and collective problem-solving~\cite{daniel_challenges_2022, fiset_effects_2024}, leading to reduced comprehension, decreased engagement, and undervaluing of contributions from colleagues~\cite{bullock_jargon_2019, brown_compensatory_2020, shulman_effects_2020}.

Strategies like preparing beforehand, asking clarifying questions during meetings, or looking up terms may alleviate some jargon-related problems; however, these approaches often fall short in real-time meeting scenarios. Preparing beforehand is often impractical because it assumes participants know precisely what terms will be challenging and have time to learn~\cite{patoko_impact_2014}. Prior research also suggests that social dynamics and hierarchical structures discourage interrupting speakers with questions, particularly among junior or culturally reserved employees~\cite{SZULANSKI20009}. Another common approach, independently searching for definitions during conversations, can introduce distractions into the conversation, which is detrimental to meetings~\cite{bailey_need_2006}.

Recent advances in speech-to-text technologies and large language models (LLMs) offer promising potential to overcome these limitations with automated, just-in-time jargon support. Prior research has explored computational techniques for jargon identification and explanation~\cite{neumann_scispacy_2019, august-etal-2022-generating, huang_understanding_2022, lucy_words_2023}, and developed augmented interfaces that enhance comprehension during meetings through interactive transcripts or captions~\cite{chandrasegaran_talktraces_2019, chen_meetscript_2023, liu_experiencing_2023, lydia2021}. However, existing systems typically neglect two critical aspects for effective knowledge support in meetings: real-time jargon support and personalization. Most prior systems either target static text content~\cite{abekawa_sidenoter_2016, head_augmenting_2021, august_paper_2023}, or they fail to consider user-specific background knowledge by providing uniform assistance to all users~\cite{liu2025exploringdesignspacerealtime}. Such generic solutions can overwhelm listeners with irrelevant or excessive information, which reduces user trust and engagement~\cite{chen_meetscript_2023, aghahoseini_investigating_2024}.

To address these gaps, we present ParseJargon,\footnote{ParseJargon stands for \underline{P}ersonalized \underline{A}ssistant for \underline{R}eal-time \underline{S}upport in \underline{E}xplaining \underline{Jargon}} a system that provides personalized jargon support in real-time online meetings. Our investigation is guided by the following research questions:


\begin{itemize}
    \item \textbf{RQ1}: How does personalization impact the effectiveness of LLM-based jargon support during online meetings?
    \item \textbf{RQ2}: What are the design and usability considerations for real-time personalized jargon-support systems?
\end{itemize}

To address RQ1, we built an initial prototype of ParseJargon for a controlled within-subjects study with seven participants, who watched each others' presentations under three conditions (7$\times$6=42 participant-presentation pairs, 14 per condition): (1) a baseline without jargon support, (2) a generic jargon support (i.e., defining the same terms for all participants), and (3) a personalized support provided by the initial prototype. We chose a simple form of personalization: first identifying and defining jargon terms, then filtering terms that the user may already know based on a one-sentence description of the listener's background. This minimal approach was meant to test an easily deployable method for personalization. Results showed that while generic support improved comprehension compared to no support, it also overwhelmed participants with excessive jargon explanations. In contrast, personalized support surfaced fewer, more relevant terms, improving comprehension and perceived value while maintaining engagement.

To address RQ2, we first used observations and participant feedback from the controlled study to derive design implications to improve the system. Refined ParseJargon incorporates a stronger meeting-profile interface and a persistent glossary, as well as more advanced personalization techniques including in-session user feedback and cross-session portable history. We then conducted system evaluations on improved personalization mechanisms and backend latency by simulating with collected controlled-study data. Simulation showed that combining in-session feedback and cross-session history further improved personalization precision from 77.51\% to 83.95\%. The latency analysis showed that ParseJargon's two-stage personalization pipeline has an acceptable real-time performance, and informed optimization directions. To complement the latency test, we made a lightweight pilot deployment of ParseJargon in a live team meeting with ten professionals. This deployment allowed us to gather preliminary user feedback on usability and identify constraints for a real-world deployment, such as caption quality.

In summary, this paper makes the following contributions:
\begin{itemize}
    \item Empirical findings from a controlled study showing that personalized jargon support, even with minimal personalization, enhances comprehension and maintains engagement compared to generic support.
    \item ParseJargon, a personalized jargon-support system for real-time online meetings, powered by LLMs to identify jargon tailored to listener's background.
    \item Design implications and usability constraints of such systems informed from user feedback, system evaluation with latency analysis, and pilot deployment.
\end{itemize}

\balance
\section{Related Work}

\subsection{Jargon Barriers and Support Technologies}

Jargon, defined as specialized terminology used within specific fields, has been frequently identified as a significant barrier in interdisciplinary communication~\cite{nickerson1999, Jeffres2011}, as experts often unconsciously utilize domain-specific language that non-experts cannot readily understand. Jargon can increase the risk of miscommunication~\cite{bullock_jargon_2019}, create resistance toward new ideas from other fields~\cite{brown_compensatory_2020}, lower engagement~\cite{shulman_effects_2020}, and intensify cognitive workload~\cite{bailey_need_2006, fiset_effects_2024}. 
These issues have been particularly difficult in workplaces for cross-team collaboration~\cite{Eppler2007, keelawat_nbguru_2023, weirup_what_2024}, interdisciplinary research~\cite{choi_multidisciplinarity_2007, wudarczyk_bringing_2021, daniel_challenges_2022}, and educational collaborations~\cite{pulimood_immersing_2024}. 

Traditional approaches to mitigating jargon have included encouraging domains experts to simplify language or employ analogies~\cite{patoko_impact_2014}, or developing domain-specific dictionaries (vocabulary lists)~\cite{gardner_new_2014}. While these strategies have been helpful, effectively adjusting one's language for new audience is notoriously difficult \citep{tullis2023curse}, requiring extensive effort and training \citep{swords2023science}. On the other side, providing static dictionaries for listeners places an onus on listeners to seek information, disrupting the natural flow of conversation.
 
Advancements in language technologies have enabled computational support for identifying and explaining jargon. Core tasks in this space include complex word identification for users~\citet{shardlow-2013-cw}, jargon complexity measured via LLMs~\cite{lucy_words_2023}, domain-specific jargon identification (for example, in biomedical text~\cite{guo_automated_2021}), and acronyms~\cite{pouran_ben_veyseh_maddog_2021}. Jargon explanation have been studied from the perspective of definition extraction~\cite{Veyseh_Dernoncourt_Dou_Nguyen_2020}, definition generation~\cite{august-etal-2022-generating}, or hybrid approaches~\cite{huang_understanding_2022}, and text simplification~\cite{martin-etal-2020-controllable, van_automets_2020}.

Building on these techniques, researchers have designed interactive systems for jargon support, particularly in asynchronous reading interfaces~\cite{semantic_reader, qlarify}.
For example, ScholarPhi~\cite{head_augmenting_2021} provides automatically generated glossaries, Paper Plain~\cite{august_paper_2023} offers in-situ definitions and plain language summaries, \citet{10.1145/3613904.3642573} augments medical progress notes, and \citet{Bao2025} supports jargon understanding via user-generated analogies. Closely related to our work, ~\citet{liu2025exploringdesignspacerealtime} explored the design space of real-time LLM-based knowledge assistance. Their findings offer valuable insights into user preferences and interface design in knowledge support when watching technical videos, including a desire for personalization. We build on this work by developing and deploying a jargon-support system for online meetings with personalization and real-time capabilities.

\subsection{Personalization in Jargon Support}

Personalization plays a critical role in tailoring jargon support to users’ individual background. Early work adapted complex word identification and lexical simplification models to individual vocabulary profiles~\cite{lee_personalizing_2018}. Subsequent efforts demonstrated that modeling word complexity at the individual level significantly improved performance~\cite{gooding_one_2022} and introduced approaches for generating personalized descriptions of scientific concepts~\cite{Murthy2021TowardsPD}. More recent research incorporated personal data into scientific jargon identification, showing that LLMs can serve as a baseline for personalized jargon support for researchers reading interdisciplinary articles~\cite{guo_personalized_2024}.

Beyond algorithmic approaches, HCI research has explored how users perceive and interact with personalized language systems, including expertise-adaptive reading interfaces for scientific information \cite{oh_news_20, tal_audience_2024} and audience-aware science journalism \cite{adar_news_17, nishal2024dejargonizingsciencejournalistsgpt4}. Recently, \citet{nishal2024dejargonizingsciencejournalistsgpt4} used LLM to help science journalists produce audience-appropriate content, revealing opportunities and challenges for dynamic adaptation; and \citet{CALISTO2025103444} showed that even perceived personalization, such as user-controlled filtering, can influence people's trust, satisfaction, and comprehension.

These studies provide the foundation for our approach, which incorporates audience background information to deliver real-time personalized jargon support. To our knowledge, our work is novel in evaluating personalized jargon support in live meetings. 

\subsection{Enhancing Meeting Communication}

The HCI community has long studied computer-mediated conversations, from text-based group chat~\cite{nam_arkose_07, do_22, do_23} to audio/video-based online meetings~\cite{dong_video_12, miller_icebreaker_21, RealityTalk}. Systems like Tilda~\cite{zhang_making_2018} and Wikum+~\cite{tian_system_2021} focus on collaborative tagging and summarizations to facilitate the sensemaking of long group chat streams. While these systems effectively support asynchronous collaboration, Meeting Bridges~\cite{wang_meeting_2024} aims to connect synchronous meetings with asynchronous conversations, helping preserve content for post-meeting engagement to reduce overload.

With increasing remote and hybrid collaboration, advances in speech-to-text techniques enables richer real-time interfaces for a variety of meeting communication tools, including visualizing ongoing meeting contents \cite{chandrasegaran_talktraces_2019}, collaborative annotation on transcripts \cite{chen_meetscript_2023}, augmenting live meeting calls \cite{Mirrorverse}, generating real-time talking points \cite{CrossTalk2023}, and distraction management \cite{Son2023}. In addition, \citet{mcgregor_meeting_17} use speech agents acting as personal assistants to boost meeting productivity, and \citet{aghahoseini_investigating_2024} leverage real-time meeting summaries to maintain engagement.

Despite these advances, most prior real-time systems have primarily focused on summarization, speaker assistance, or distraction management, leaving jargon-related comprehension challenges relatively unexplored. Our system targets this gap by leveraging real-time meeting transcripts to address jargon barriers during online meetings, with a focus on how personalization affects comprehension and engagement in cross-background conversations.

\section{Controlled Study of the Initial Prototype}
\label{sec:controlled_study}

To probe whether personalization can meaningfully improve jargon support in meetings, we first built an initial prototype of ParseJargon that uses a minimal personalization technique: filtering identified terms using a listener's one-sentence profile. We then evaluated this prototype in a controlled within-subjects study. Beyond testing effectiveness, this study also served a formative role to inform design improvements for a refined system described in Section~\ref{sec:system}.

\subsection{Initial Prototype of ParseJargon}
\label{subsec:initial_prototype}

\begin{figure*}[h]
  \centering
  \begin{subfigure}[t]{0.49\textwidth}
    \centering
    \includegraphics[width=1.02\textwidth]{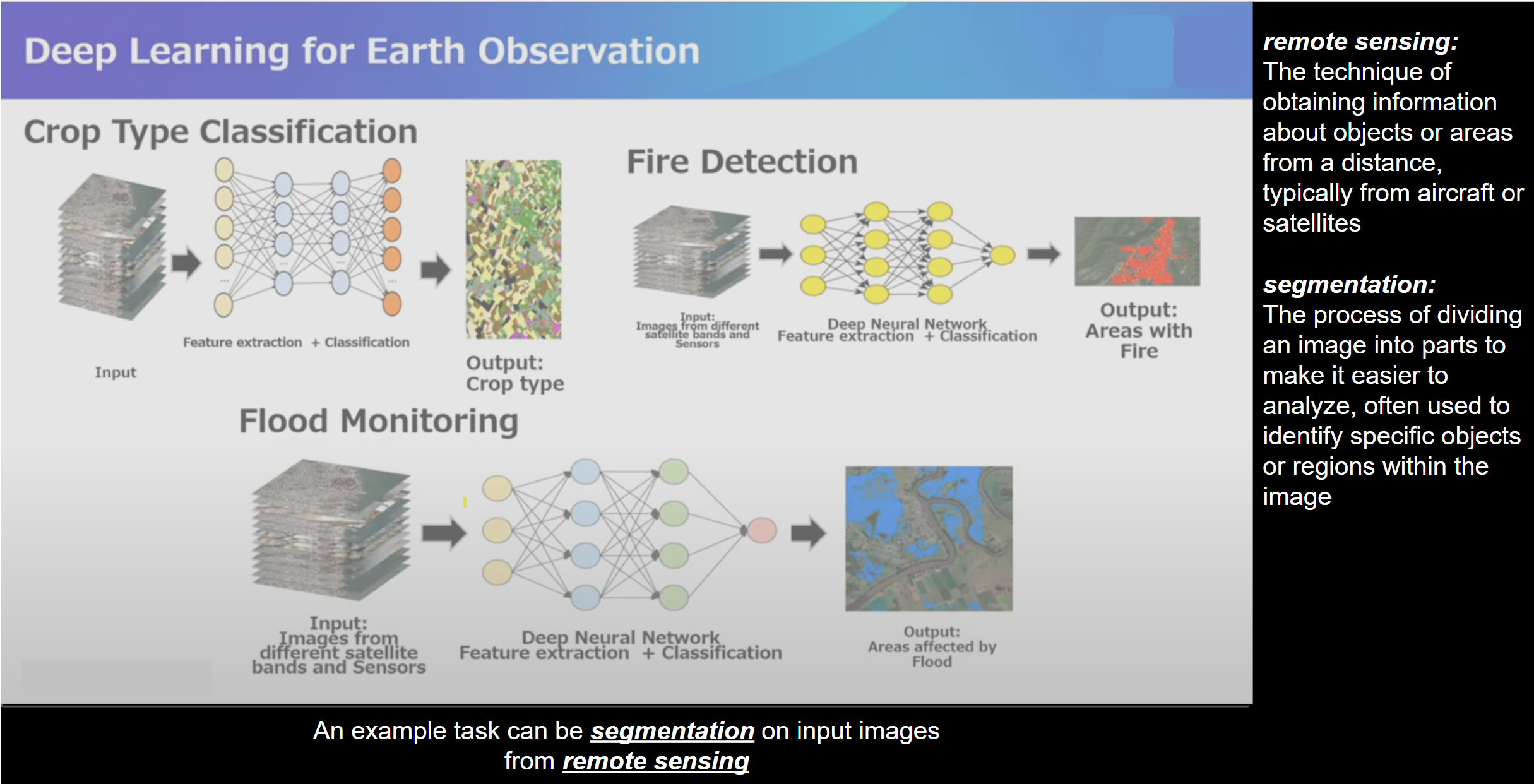}
    \caption{Generic Jargon (Condition 1)}
    \label{fig:general_interface}
  \end{subfigure}
  \hfill
  \begin{subfigure}[t]{0.49\textwidth}
    \centering
    \includegraphics[width=1.02\textwidth]{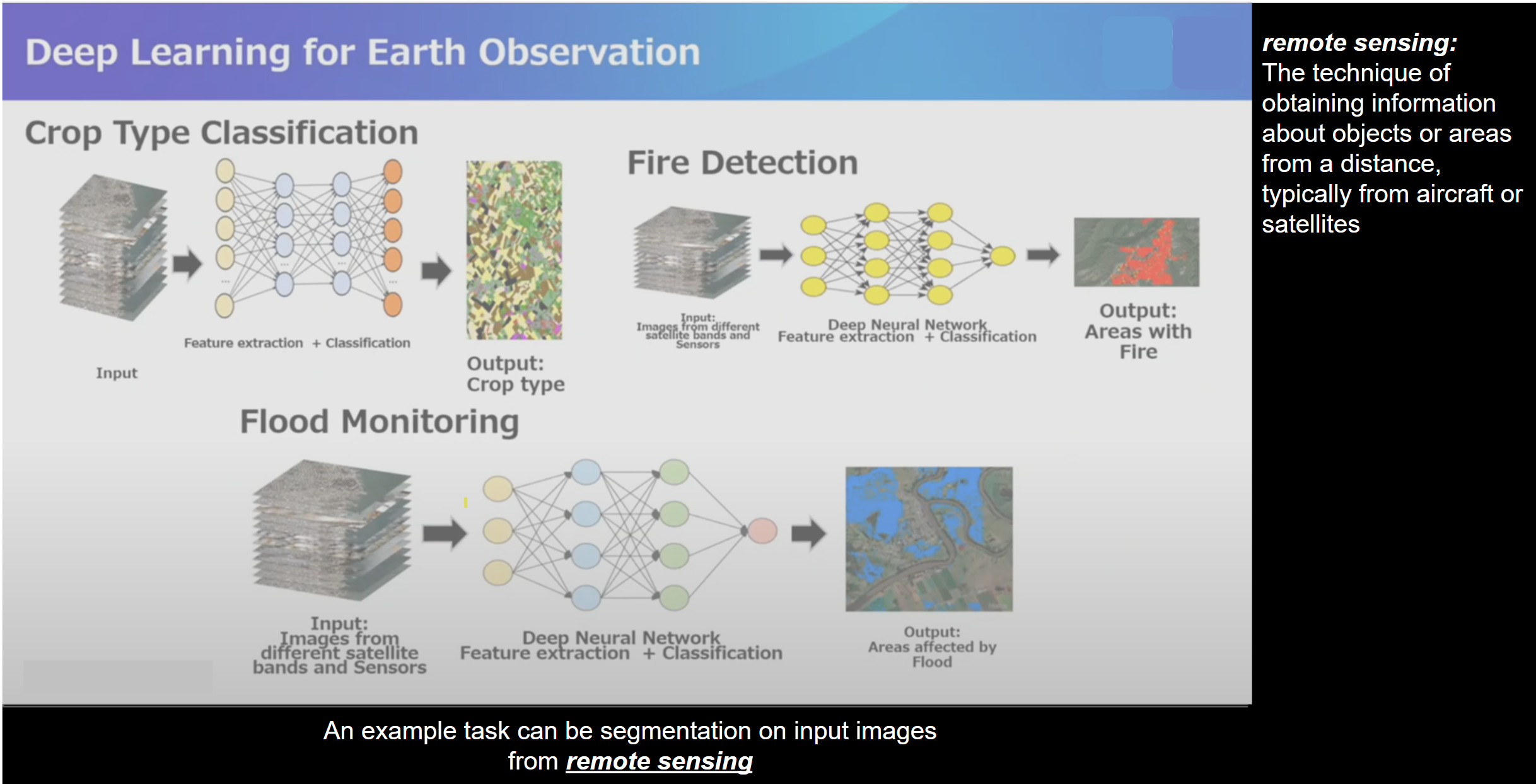}
    \caption{Personalized Jargon (Condition 2)}
    \label{fig:personalized_interface}
  \end{subfigure}
  \caption{Generic vs.\ Personalized: (a) shows generic jargon support that assumes the listener has no expertise of any domains; (b) shows personalized jargon support in a scenario that the listener provides a profile indicating background in computer vision but not earth science, so \textit{\textbf{segmentation}} is filtered out and only \textit{\textbf{remote sensing}} is displayed.}
  \label{fig:interfaces}
\end{figure*}

In live use, the initial prototype takes meeting speech captions, generated by online meeting platforms' speech-to-text engines, as input and outputs a glossary of identified jargon terms with definitions. Because meeting-platform transcriptions are not always reliable~\cite{picovoice}, we used \textbf{pre-recorded videos} in this controlled study and fed the prototype \textbf{manually verified transcripts}. This allowed us to evaluate personalized jargon support as a proof-of-concept while isolating the effect of personalization from transcription inaccuracies or latency issues.

The prototype leverages an LLM\footnote{Used \texttt{gpt-4o-2024-05-13}, same in system evaluation and deployment study} to perform three interconnected tasks in a pipeline: \textit{jargon identification}, \textit{jargon explanation}, and \textit{personalization}. These tasks are executed through prompting, where all prompts are provided in Appendix~\ref{app:prompt}.

\begin{itemize}
\item \textbf{Jargon Identification \& Explanation:} The prototype first fetches the transcript generated by the service provided in online meeting platforms, then prompts an LLM to identify potential jargon in the transcript and generate concise plain-language definitions for each term. This process uses a single combined prompt, analyzing each sentence of the meeting transcript sequentially. Each identified jargon term is defined only once.
\item \textbf{Personalized Filtering:}
To tailor jargon support to individual audience members' expertise, we apply a second filtering step. A second prompt takes a user profile and removes identified terms that the user likely already knows. In our experiment, we test having the participant (i.e., the listener) enter a one-sentence description of their educational background and/or job role as their profile (e.g., “\textit{I am a Physics PhD, working as a research intern in the Quantum Computing team}”).
\end{itemize}

We use a two-stage pipeline (identify+define, then filter) to isolate the effect of personalization on term selection, rather than explanation generation. The identified jargon terms with definitions are then displayed in real time with a simple glossary interface, implemented as a sidebar next to the main meeting/video screen. Figure~\ref{fig:interfaces} shows the controlled-study interface under the generic and personalized conditions (i.e., without vs. with the personalized filtering). This brief example illustrates how a listener's profile can help reduce false‑positive jargon terms that the listener likely already knows.

Each term in the sidebar is displayed until the next term is identified with a minimum display of 7 seconds, set based on average reading speed~\cite{BRYSBAERT2019104047} and an internal pilot. If a new jargon term is identified before the minimum display of an existing definition, both will appear and the first term will be replaced after 7 seconds, informed by the same mechanism from prior research~\cite{liu2025exploringdesignspacerealtime}. Identified terms are also highlighted in the caption for easy recognition.


\subsection{Methods}
\label{subsec:controlled_methods} 

\subsubsection{Participants}

We recruited seven participants in a technology company through emails and in-person outreach. All participants are interns who worked on different projects from different teams to minimize their prior knowledge about each other's work. Demographics information, including their education and job role to create the one-sentence profile, is included in Appendix Table~\ref{tab:participants}.

\subsubsection{Experimental Conditions}

Each participant prepared a 10-minute presentation with slides about their ongoing project. Presentations were recorded and transcribed via Microsoft Teams, the company’s primary communication platform. All transcripts were manually verified for correctness. We created three experimental conditions using the recorded and transcribed presentations:

\begin{itemize}
    \item \textbf{Generic Support (Condition 1):} Recordings were processed by the initial prototype without profile-based filtering (Figure~\ref{fig:general_interface}).
    \item \textbf{Personalized Support (Condition 2):} Recordings were processed by the prototype with profile-based filtering enabled (Figure~\ref{fig:personalized_interface}).
    \item \textbf{No-support Baseline (Condition 0):} Original recordings with no jargon support; cleaned transcripts were displayed to match Conditions~1 and~2 but no jargon highlighting.
\end{itemize}

\subsubsection{Procedure}
We employed a within‑subjects design where each participant viewed all presentations from the other six participants. Participants watched recordings individually in two sessions (\(\sim\)45 minutes each), viewing three presentations per session (one per condition). Sessions were spaced at least one day apart to minimize fatigue. We approximated a Latin‑square design by rotating the three conditions so counts were roughly evenly distributed across first, middle, and last positions. This yielded 14 unique viewing experiences per condition and 42 participant‑presentation pairs (a full schedule of video and condition ordering per participant is in Appendix Table~\ref{tab:counterbalance}). 

Participants watched on laptops without pausing or navigating but were free to consult external resources (e.g., web search, LLM queries) to approximate live meeting behavior. The researcher observed screens to verify adherence. Brief follow-up interviews were conducted at the end of the second session to contextualize quantitative results and surface limitations of the minimal prototype that informed later design refinements.

\subsubsection{Metrics}

\paragraph{Glossary helpfulness rate.}
After Conditions~1 and~2, participants reviewed the glossary (all identified jargon terms) and labeled each term as "helpful" or "not helpful". We computed \textit{helpfulness rate} as the proportion labeled helpful, which can be interpreted as the \textit{precision} of jargon identification.

\paragraph{Self‑reported measures.}
After each presentation, participants completed a short survey with the following 5‑point Likert-style questions (1 = not at all, 5 = very):
\begin{itemize}
    \item \textit{Comprehension Confidence}: “How confident do you feel in your understanding of the presentation?”
    \item \textit{Engagement}: “How engaged were you while following the presentation?”
    \item \textit{Perceived Value}: “How valuable do you think the presented work is?”
    \item \textit{Glossary Usefulness} (only Conditions~1\&2): “How useful were the term explanations in the glossary sidebar?”
\end{itemize}

\paragraph{Comprehension assessment.}
To capture how effectively participants grasped and engaged with the ideas in the presentation, we had participants write a one-sentence takeaway and a one-sentence question for each presentation. Original presenters anonymously evaluated these for \textit{clarity}, \textit{relevance}, and \textit{depth} (5‑point Likert scales, 1 = least, 5 = most). This measure focuses on the speaker’s evaluation of listener comprehension, aligning with workplace communication needs~\cite{Eppler2007, weirup_what_2024}.

\subsubsection{Analysis}
\label{subsec:analysis}
We report the statistical comparisons on the Likert-scale metrics between conditions using Wilcoxon signed-rank test. We chose this non-parametric approach given the ordinal nature of Likert-scale data and the relatively small participant sample size~\cite{mircioiu_comparison_2017}. Effect sizes were calculated. Holm–Bonferroni corrections were applied to control for Type~I errors due to multiple comparisons~\cite{aickin_adjusting_1996}. We report significance based on corrected p-values.

\subsection{Findings}
\label{subsec:controlled_findings} 

Table~\ref{tab:helpful_rate} compares the number of jargon terms identified and helpfulness rate between generic and personalized condition, and Table~\ref{tab:outcomes-wide} summarizes both self-reported measures and presenter-graded comprehension. Full statistics are provided in Appendix Table~\ref{tab:full-stats-selfreport}.

\begin{table}[h]
\small
  \caption{Average number of identified terms and helpfulness rate per participant-presentation.}
  \begin{tabular}{l|l||l}
    \toprule
    \textbf{Condition}&\textbf{Helpful / Total Terms} & \textbf{Helpfulness Rate} \\
    \midrule
    Generic      & 10.29 / 22.57 & 47.03\% \\
    Personalized & 7.64 / 9.71  & 77.51\% \\
    \bottomrule
  \end{tabular}
  \label{tab:helpful_rate}
\end{table}

\begin{table*}[h]
\small
  \caption{Self-reported measures and presenter-graded comprehension by condition (mean $\pm$ SD; 1--5 Likert). 
  Self-report metrics include \textit{comprehension confidence}, \textit{engagement}, \textit{perceived value}, and \textit{glossary usefulness}, and presenter-graded comprehension metrics (aggregated across takeaways+questions) include \textit{clarity}, \textit{relevance}, and \textit{depth}. Statistical significance ($^\star$\,vs.\ Baseline; $^\dag$\,vs.\ Generic) indicates corrected $p{<}.05$ using Wilcoxon signed-rank tests.}
  \label{tab:outcomes-wide}
  \centering
  \setlength\tabcolsep{4pt}
  \begin{tabular}{l|cccc|ccc}
    \toprule
    Condition & Comprehension & Engagement & Value & Usefulness & Clarity & Relevance & Depth \\
    \midrule
    Baseline
      & 3.07$\pm$0.62 & 3.93$\pm$0.83 & 3.57$\pm$0.65 & -- 
      & 4.18$\pm$0.80 & 3.64$\pm$1.06 & 3.25$\pm$1.17 \\
    Generic
      & 3.79$\pm$0.70$^\star$ & 3.64$\pm$1.01 & 3.93$\pm$0.62 & 3.93$\pm$0.83
      & 4.21$\pm$0.73 & 4.43$\pm$0.65$^\star$ & 3.68$\pm$0.72 \\
    Personalized
      & 4.29$\pm$0.61$^\star$$^\dag$ & 4.29$\pm$0.73 & 4.43$\pm$0.51$^\star$$^\dag$ & 4.64$\pm$0.50$^\dag$
      & 4.36$\pm$0.63 & 4.39$\pm$0.45$^\star$ & 4.04$\pm$0.57$^\star$ \\
    \bottomrule
  \end{tabular}
\end{table*}

\paragraph{Personalized jargon support identifies fewer terms and is more precise than generic support.} 

From Table~\ref{tab:helpful_rate}, personalization reduced glossary size by ~57\% (from 22.6 to 9.7 terms) while improving helpfulness rate from 47\% to 77.5\%. Showing fewer but more relevant entries can limit on-screen distractions during presentation flow, helping participants stay with the speaker rather than triaging definitions. This increased precision in the personalized condition was associated with higher self-reported comprehension and perceived value compared to generic support (Table~\ref{tab:outcomes-wide}).

\paragraph{Personalized support improved self-reported comprehension and perceived value while maintaining engagement.}

As shown in Table~\ref{tab:outcomes-wide}, both the generic and personalized conditions increased self-reported comprehension compared to the baseline, with the personalized condition showing greater improvement. While both experimental conditions improved participants' perceived value of others' presented work, only the personalized condition showed significance in rating improvement to the baseline condition. 
However, participants reported lower engagement in the generic condition than baseline, whereas personalized support maintained engagement with a higher score than both baseline and generic support. Interviews suggested that participants felt overwhelmed by the excessive number of definitions (6 participants) in the generic condition. \textit{"Too many term definitions with very short reading time"}, as described by P1. Some participants even described this as \textit{"annoying"} (P4) or even \textit{"offensive (because) the system treats me like I know nothing"} (P3). 

\paragraph{Jargon support led to higher scores in comprehension assessment.} 

Beyond increased self-reported comprehension confidence, presenters rated listeners' one-sentence takeaways and questions as significantly more relevant under both jargon-support conditions, and with greater depth under personalized support (Table~\ref{tab:outcomes-wide}). Clarity remained relatively unaffected, likely due to the fact that clarity depends more on participants' writing skills than on their presentation understandings. Qualitatively, both the takeaways and questions stayed relatively vague and non-specific in the baseline condition. The ones from the generic condition were more targeted to key topics, while the personalized-condition ones also included in-depth technical details. As an example, we include all takeaways and questions from two presentations in Appendix Table~\ref{tab:comprehension_example}. 

\paragraph{When minimal personalization doesn't work well.} 
While personalized support was generally beneficial, we observed cases where minimal personalization with one-sentence profile performed sub-optimally. For instance, in P1's presentation, many jargon terms were related to business operations. Although P2 (the listener) was majoring in mathematics, they had start-up business experience that was not included in the one-sentence profile, so the business terms remained unfiltered. This highlights how quality and completeness of user profiles influence personalization precision. 


\subsection{Design Implications}
\label{subsec:design_implications}

Taken together, the controlled study shows that even minimal profile-based personalization can make jargon support substantially more helpful by improving glossary precision and reducing information load. However, failure cases and post-study interviews also revealed limitations of the initial prototype and suggest three design implications that motivate the refined ParseJargon system in Section~\ref{sec:system}.

\paragraph{Capture expertise with lightweight but richer profiles.}
A single free-text sentence kept onboarding simple, but it often omitted relevant expertise that affected filtering precision. For example, in the failure case described in Section~\ref{subsec:controlled_findings}, business-related terms remained surfaced because the listener's prior startup experience was not reflected in their one-sentence profile. Rather than requiring an extensive user model, though, a refined system can collect expertise in a lightweight but more structured way, such as separating professional and educational background with explicit keywords of familiar domains or topics.

\paragraph{Support in-session user feedback to improve personalization.}
The initial prototype offered no way to respond when a surfaced jargon term was actually unnecessary. Participants wanted to tell the system that they were already familiar with certain terms, rather than being locked into the initial profile. For instance, P7 noted in the post-interview that they had taken a quantum computing course that was not captured in their initial profile and wished they could indicate that they did not need support for quantum-related terms once such terms first appeared, so that similar terms would not continue to be explained for the remainder of the meeting. A refined system should support low-effort per-term user feedback during the meeting to further improve personalization by distinguishing surfaced terms that were helpful from those already known.

\paragraph{Preserve surfaced terms within and across meetings.}
Four participants also wanted to revisit surfaced definitions when reading time was limited. As P5 noted, \textit{"There are too many definitions... I haven't finished reading and it disappeared... I cannot go back and check (the definition)."} A refined system can therefore preserve surfaced terms in a persistent glossary during the meeting. In addition, the user's authored profile and interaction history---including these surfaced terms and listener feedback---can be saved in a portable form for reuse across meeting sessions. Users should remain in control of whether this prior usage history is carried forward, but when reused, it can bootstrap personalization in later meetings.

\section{Refined ParseJargon}
\label{sec:system}

\begin{figure*}[h]
  \centering
  \includegraphics[width=0.8\linewidth]{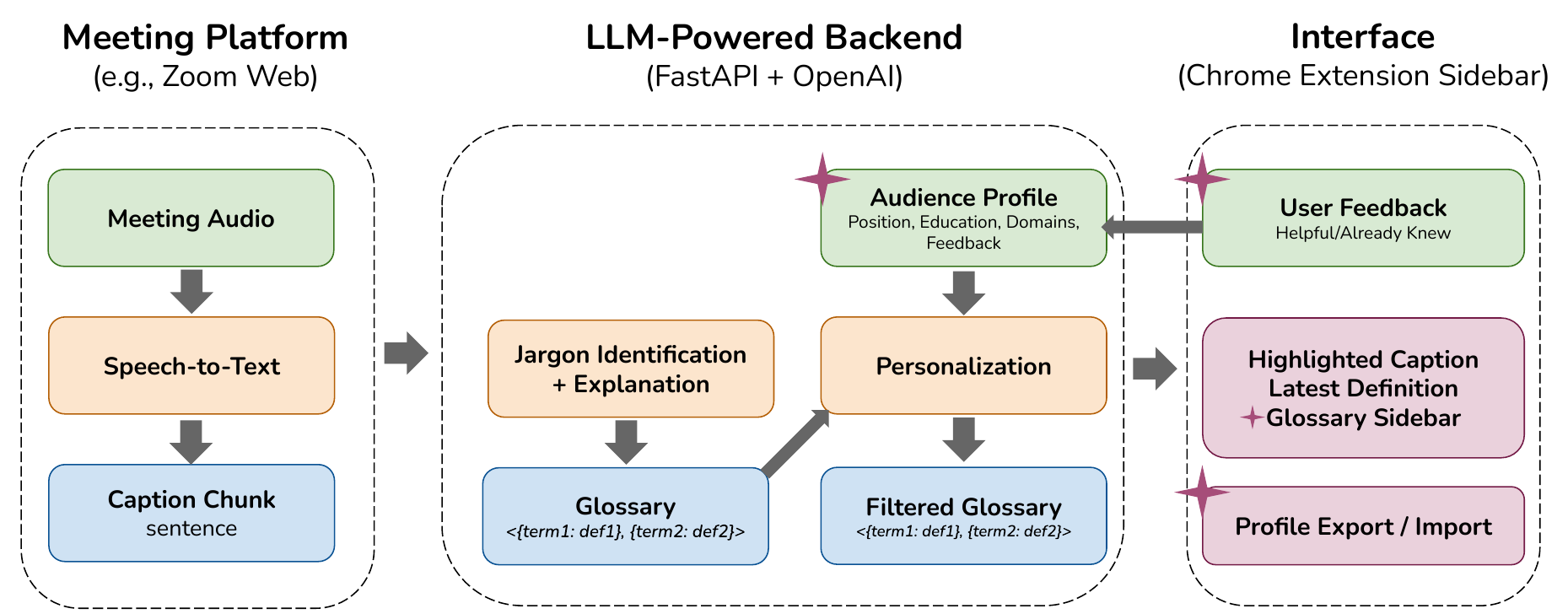}
  \caption{ParseJargon architecture. The system retains the core pipeline from the initial prototype: speech-to-text, jargon identification and explanation, and personalized filtering, and extends it with a stronger meeting-profile interface, per-term feedback during meetings, and portable cross-session profile reuse (purple star indicating updates in the refined system).}
  \label{fig:architecture}
\end{figure*}

Guided by the design implications from Section~\ref{subsec:design_implications}, we refined ParseJargon from its initial study prototype version into a system for real-time, personalized meeting support. The refined system retains the core two-stage pipeline introduced in Section~\ref{subsec:initial_prototype} (jargon identification with explanation followed by personalized filtering), but augments it with additional personalization techniques and interface enhancements. 

\subsection{Overview}
\label{subsec:overview}

Figure~\ref{fig:architecture} shows the refined system architecture. As in the initial prototype, meeting speech is transcribed by the meeting platform and processed sentence by sentence. The backend first identifies candidate jargon terms and generates concise definitions, then applies a personalization step before surfacing terms in the interface. The core prompt-based pipeline remains unchanged from Section~\ref{subsec:initial_prototype}; the main refinement is that personalization now conditions on multiple lightweight sources: an explicit meeting profile authored by the user, feedback collected during the current meeting, and an optional portable profile imported from prior meetings. The prompt used in the personalization stage is included in Appendix~\ref{app:prompt}

As shown in Figure~\ref{fig:new_system}, the interface is correspondingly refined into two coordinated parts: a meeting-profile panel (Figure~\ref{fig:new_system} (3)) used to enter or import profile information before a meeting session, and an in-session sidebar that separates the latest term from a persistent glossary (Figure~\ref{fig:new_system} (2)). This organization supports both glanceable real-time assistance and later revisiting of earlier terms when reading time is limited. Since the core backend and term display logic is already described in Section~\ref{subsec:initial_prototype}, we focus below on the added and improved personalization and interface mechanisms.

\subsection{Refined Personalization and Interface}
\label{subsec:refined_features}

This subsection describes the three refinements directly informed by the design implications in Section~\ref{subsec:design_implications}. These refinements do not change the core jargon-detection pipeline; instead, they make personalization easier to specify, easier to correct, and easier to reuse across meetings.

\subsubsection{Richer Meeting Profile}
\label{subsubsec:meeting_profile}

The initial prototype relied on a single free-text sentence to describe the listener's background. While simple, this format often omitted relevant expertise that affected filtering precision. The refined meeting-profile interface therefore collects background information through several lightweight fields, such as a user's current job position, education level, and familiar domains. Familiar domains can be entered as custom keywords or selected from suggested options, allowing users to quickly represent multi-dimensional expertise without writing a long narrative. Importantly, the profile remains user-authored, which keeps personalization inspectable and gives users explicit control over how their expertise is represented to the system.

\subsubsection{In-Session Per-Term Feedback}
\label{subsubsec:feedback}

Unlike the initial prototype, the refined system supports in-session per-term user feedback on surfaced terms. For each term, users can indicate either \textit{Helpful} or \textit{Already Knew}. \textit{Helpful} records that the surfaced term and its explanation met the current listener's needs, while \textit{Already Knew} signals that the term was unnecessarily surfaced for this user, indicating that a correction should be applied to personalization. In our current implementation, these feedback signals are stored as session state and appended as context to subsequent personalization prompts (Appendix~\ref{app:prompt}). This feedback design keeps interaction lightweight enough for live meetings while distinguishing between "terms that helped me" and "terms that should not have been shown to me."

\subsubsection{Persistent Glossary with Portable Cross-Session Reuse}
\label{subsubsec:portable_profile}

The refined interface separates a latest-term card for glanceable support from a persistent glossary that accumulates all jargon terms surfaced during the meeting. This addresses a key limitation of the initial study interface: listeners may not have enough time to fully read a term when it first appears, or they may want to revisit terms later in the meeting. By keeping prior terms accessible, the persistent glossary allows users to check earlier definitions without losing the flow of the meeting. The glossary interface also supports searching, filtering, and deleting individual term entries. At the end of a meeting session, users can export their current meeting profile together with the glossary, including surfaced terms and associated feedback, as a \texttt{JSON} file. In a later meeting, users can import this file to pre-populate their profile and optionally reuse prior glossary history for personalization. This exported file serves as a simple portable profile that can be carried across meetings without building a centralized account system.

\begin{figure*}[h]
  \centering
  \includegraphics[width=0.9\textwidth]{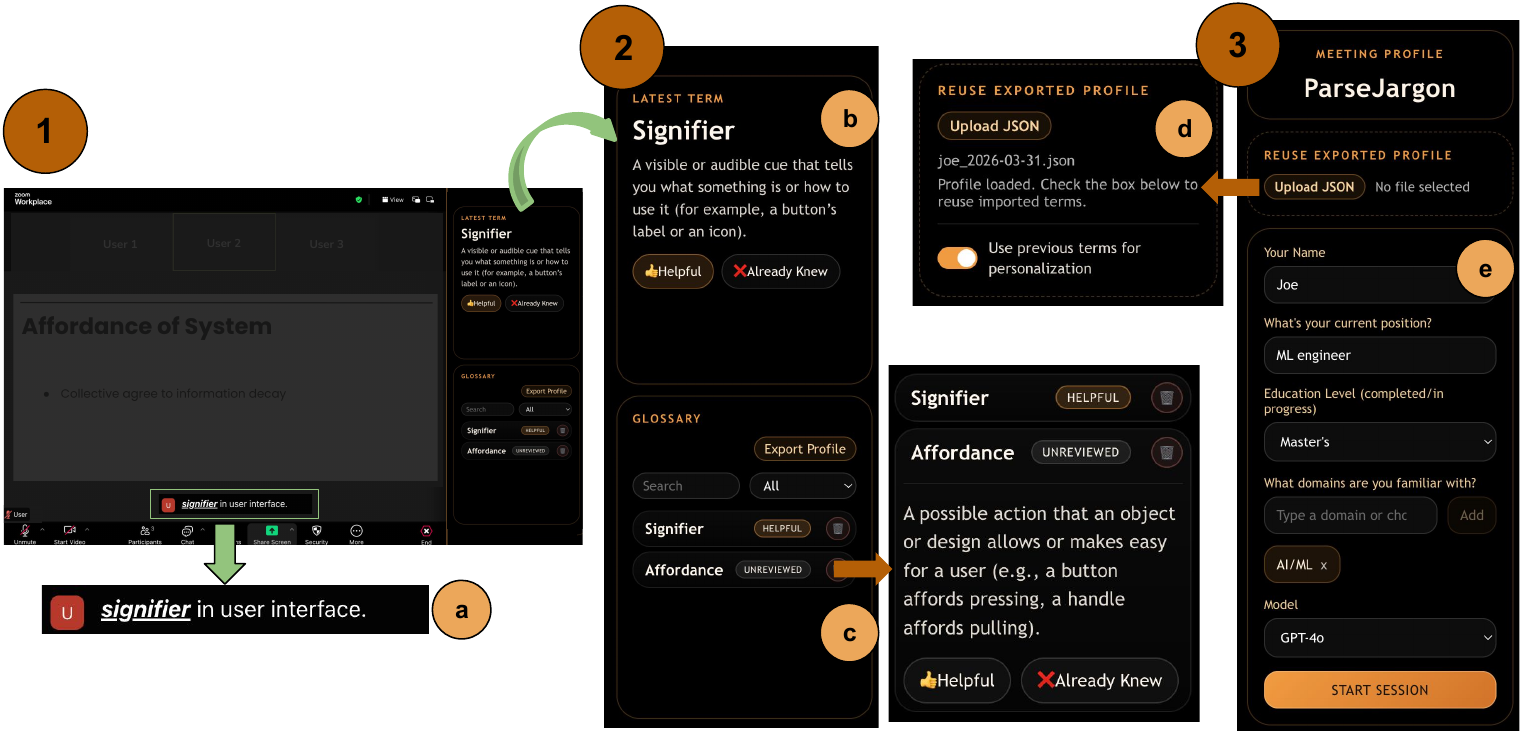}
  \caption{The ParseJargon sidebar interface embedded in Zoom Web: (1) Overview of how ParseJargon is embedded and looks like in Zoom Web; (2) In-session sidebar during a meeting; and (3) User profile sidebar before starting a meeting session. \newline Interface components include: (a) Highlights on the jargon terms in real-time caption; (b) The Latest Term panel displays real-time jargon detected from live captions, with inline feedback buttons (Helpful and Already Knew) to refine personalization. (c) The Glossary panel accumulates all identified terms for the session, supporting search, status filtering, term deletion, and one-click profile export for reuse in future meetings. (d) Users can upload a previously exported profile to auto-fill personalization fields and optionally reuse prior terms via toggle switches. (e) User profile form collects user information including name, job position, education level, and familiar domains.}
  \label{fig:new_system}
\end{figure*}

\subsection{Implementation Details}
\label{subsec:implementation}

ParseJargon is implemented as a Chrome extension integrated into the web version of Zoom. The frontend is built with \texttt{React.js}, while a Python \texttt{FastAPI} backend manages calls to OpenAI API and maintains per-session state. In the following system evaluation and deployment study, we use \texttt{gpt-4o} with the server deployed on \texttt{Heroku} and connected to a \texttt{PostgreSQL} database. User profile and glossary can be exported and reused through local \texttt{JSON} files.
\section{System Evaluation}
\label{sec:system_evaluation}

To evaluate the refined system, we reused the data collected in the controlled study to simulate ParseJargon on manually verified transcripts and participant-validated glossary labels. The available data consist of one-sentence profiles and post-hoc \textit{helpful} vs.\ \textit{not helpful} judgments, which allow us to approximate the in-session user feedback and cross-session portable glossary introduced in Section~\ref{subsec:refined_features}. We then report a latency analysis of the core personalized pipeline, compared against a generic baseline, using the same simulation setup.

\subsection{Simulation Setup}
\label{subsec:simulation_setup}

We used the 14 participant-presentation pairs from the personalized condition in Section~\ref{sec:controlled_study}. For each session, participants had already reviewed all surfaced jargon terms and labeled them as \textit{helpful} or \textit{not helpful}. To simulate the refined system's in-session feedback mechanism, we treat terms labeled \textit{helpful} as if the user clicked \textit{Helpful}, and terms labeled \textit{not helpful} as if the user clicked \textit{Already Knew}. These simulated signals are then appended to subsequent personalization prompts within the same session.

We also simulate cross-session reuse. Since each participant experienced two personalized sessions, we use one personalized session as exported history to warm-start the same participant's subsequent personalized session, and then reverse the direction. This yields 14 cross-session simulations in total. The imported history contains the prior session's surfaced jargon terms together with their associated feedback.

This simulation should be interpreted as an approximation rather than a direct measure of live longitudinal use. In particular, a term labeled \textit{not helpful} may reflect multiple causes beyond prior familiarity, and each participant contributes only two personalized sessions for cross-session reuse. Still, this setup offers a practical estimate of how the refined personalization mechanisms can improve jargon selection using the gold-standard labels already collected in the controlled study. The same transcript corpus also provides a testbed for our latency analysis.

\subsection{Personalization Evaluation}
\label{subsec:personalization_eval}

\begin{table}[h]
\small
  \centering
  \caption{Simulated personalization performance on the 14 personalized participant-presentation pairs.}
  \label{tab:simulation_personalization}
  \begin{tabular}{l|l|l}
    \toprule
    \textbf{Configuration} & \textbf{Helpful / Total Terms} & \textbf{Helpfulness Rate} \\
    \midrule
    Static Profile & 7.64 / 9.71 & 77.51\% \\
    + In-Session Feedback & 6.94 / 8.32 & 83.41\% \\
    + Cross-Session Reuse & 7.36 / 9.11 & 80.79\% \\
    + Feedback \& Reuse & 6.85 / 8.16 & 83.95\% \\
    \bottomrule
  \end{tabular}
\end{table}

Table~\ref{tab:simulation_personalization} summarizes the simulated personalization performance. We compare the original static-profile personalized condition from the controlled study (Table~\ref{tab:helpful_rate}) with three refined variants: in-session feedback, cross-session glossary reuse, and the combination of both.

Both refined mechanisms improve over the static-profile baseline, but in-session feedback is more impactful. Simulated in-session feedback increased the helpfulness rate from 77.51\% to 83.41\%, suggesting that feedback can suppress redundant terms that static profiles miss without requiring a more extensive user model. Cross-session reuse alone also improved performance to 80.79\%, indicating that portable glossary history can warm-start personalization across meetings, even when the two presentations are on different projects with only partial overlap in topics. Combining cross-session reuse with in-session feedback achieved the highest helpfulness rate at 83.95\%, but only marginally improved over in-session feedback alone. This suggests that the benefit of these lightweight personalization mechanisms begins to saturate under the current simulation setup, where each user contributes only two personalized sessions and cross-session memory is represented by simple glossary history. 

Overall, these results suggest that both mechanisms are helpful, but in-session feedback is the more straightforward and impactful improvement in the current system. At the same time, the positive effect of cross-session reuse suggests that ParseJargon has the potential to become more effective as a longer-term personal assistant that accumulates richer evidence about a user's knowledge and preferences over many meetings.

\subsection{Latency Analysis}
\label{subsec:latency_eval}

We also measure ParseJargon's backend latency as an indicator of real-time capability using the same simulation setup. We report \textit{pipeline latency per caption chunk}: the elapsed time from when the transcription service prepares a caption chunk to be displayed to when the system's response (i.e., any definitions) for that chunk is returned. We evaluated two conditions: (1) a generic baseline, as in the controlled study, that includes jargon identification and explanation without the personalization filter; and (2) a personalized condition using the core two-stage identify–then–filter pipeline.

There were 787 caption chunks across the seven recordings. In the personalized condition, each transcript was measured twice with two participants' profiles, yielding 1574 chunks. Because latency varies with API traffic and network conditions, we ran the test twice at different days and times under the same network environment (800 Mbps Wi-Fi), resulting in 1574 chunks for the generic condition and 3148 chunks for the personalized condition. Based on the two runs, we report medians (50th percentile) and tails (95th percentile), as tail latency most strongly affects perceived responsiveness in interactive services~\cite{10.1145/3232559}. We did not measure end-to-end UI latency (from meeting-platform caption to on-screen glossary rendering), which depends on device factors beyond our control.

To capture the timeliness of jargon presentation, we compare latency with the length of time that caption chunks were displayed on screen (\textit{chunk duration}): caption chunks remained visible for a median of 4.15\,s and a 95th percentile of 6.62\,s. As shown in Table~\ref{tab:latency_subrows}, the median and tail of backend latency were 0.32/1.36\,s (generic) and 0.42/1.73\,s (personalized). While personalization introduces a modestly higher latency, these values remained well below \textit{chunk duration}, suggesting that definitions arrived well before the captions containing referenced terms disappeared.

\begin{table}[h]
\small
  \centering
  \caption{Pipeline latency (seconds) per caption chunk. Median and tail (95th percentile) for all chunks and subsets stratified by whether the chunk emitted any jargon terms.}
  \label{tab:latency_subrows}
  \begin{tabular}{l l l l l}
    \toprule
    \textbf{Condition} & \textbf{Subset} & \textbf{N} & \textbf{Median} & \textbf{Tail} \\
    \midrule
    Generic & All chunks & 1574 & 0.32 & 1.36 \\
            & \quad No jargon & 1350 & 0.31 & 0.53 \\
            & \quad $\geq$1 terms & 224 & 1.11 & 2.25 \\
    Personalized & All chunks & 3148 & 0.42 & 1.73 \\
            & \quad No jargon & 2980 & 0.41 & 1.35 \\
            & \quad $\geq$1 terms & 168 & 1.60 & 3.45 \\
    \bottomrule
  \end{tabular}
\end{table}

Latency is especially higher in the personalized condition on chunks that emit terms due to the process of first generating definitions then filtering. This suggests that the main source of overhead is the current two-stage design. While latency of ParseJargon remains acceptable for real-time use, future versions could reduce this gap by merging jargon identification and personalized filtering into a single prompt through one API call, and by using optimization strategies such as batching and caching.
\section{Pilot Deployment}
\label{sec:deployment}

To complement the controlled study and the system evaluation, we conducted a lightweight pilot deployment of ParseJargon in a real-time team meeting. This deployment was not designed to measure the effectiveness of personalized jargon support or separately evaluate each system component. Instead, it aimed to assess ParseJargon's practical usability, perceived cognitive workload, and perceived usefulness in real-world live meetings.

\subsection{Methods}

\subsubsection{Participants}
We deployed ParseJargon during a regularly scheduled weekly team meeting with ten members: one director, six senior researchers, and three junior researchers.
The meeting was at the same technology company as the controlled study, but with no participant overlap. The meeting involved junior researchers presenting project updates. Although team members shared broad organizational context, differences in specific project and research domains still introduced unfamiliar terminology.

\subsubsection{Procedure}
Participants installed ParseJargon as a Chrome extension following brief instructions, after which we introduced its key features ($\approx$10 minutes). Participants then joined the Zoom meeting via their browser and entered their profile. The meeting proceeded normally while ParseJargon surfaced jargon definitions and maintained a persistent glossary. The session lasted approximately 45 minutes, with three 10-minute presentations followed by short discussions. Immediately after the meeting, participants completed an online survey. Because this was a single live deployment without a comparison condition, the goal was to assess whether ParseJargon could be used without introducing noticeable disruption, rather than to estimate comparative effects.

\subsubsection{Measures}
The post-meeting survey assessed the system along four dimensions:
\begin{itemize}
    \item \textit{Usability}: ease of use, feature integration, and willingness for future use (5-point Likert items, 1 = least, 5 = most),
    \item \textit{Cognitive workload}: NASA-TLX~\cite{hart1988development}, dropping the \textit{Performance} dimension to focus on the other five workload dimensions (21-point scale, 1 = lowest load, 21 = highest load),
    \item \textit{Perceived usefulness}: how ParseJargon affected comprehension, engagement, and perceived value of colleagues' presentations (5-point Likert items. 1 = not at all, 5 = very, questions taken from Section~\ref{subsec:controlled_methods}),
    \item \textit{Qualitative feedback} on overall experience, strengths, limitations, and suggestions.
\end{itemize}

\subsection{Findings}
\label{subsec:deployment_results}

Overall, the pilot deployment suggests that ParseJargon can be integrated into live meetings with low cognitive overhead, while still experiencing practical limitations around caption quality and meeting genre.

\begin{table}[h]
\small
\centering
\caption{NASA-TLX results from the pilot deployment (out of 21; lower indicates lower load).}
\label{tab:nasa_tlx_results}
\begin{tabular}{l c}
\toprule
\textbf{Dimension} & \textbf{Score (Mean $\pm$ SD)} \\
\midrule
Mental Demand & 4.8 $\pm$ 3.39 \\
Physical Demand & 2.3 $\pm$ 1.83 \\
Temporal Demand & 5.2 $\pm$ 4.54 \\
Frustration Level & 5.3 $\pm$ 3.65 \\
Effort & 5.1 $\pm$ 4.25 \\
\bottomrule
\end{tabular}
\end{table}

\subsubsection{Usability and Cognitive Workload}
Participants rated ParseJargon's usability positively, with mean scores of 4.3 (SD $= 0.82$) for ease of use, 4.5 (SD $= 0.53$) for feature integration, and 3.9 (SD $= 1.1$) for willingness for future use. NASA-TLX scores likewise indicate low workload across dimensions (Table~\ref{tab:nasa_tlx_results}), suggesting that the system did not substantially interfere with normal meeting participation. Participants described the system as \textit{"a smooth interface and easy integration with Zoom-like applications"} (P4) and \textit{"automatic and non-intrusive"} (P10).

\subsubsection{Perceived Usefulness}
Participants also reported positive perceived effects on comprehension (mean $= 3.9$, SD $= 1.37$), engagement (mean $= 3.6$, SD $= 1.35$), and perceived value of colleagues' work (mean $= 4.0$, SD $= 1.05$), consistent with the trends observed in the controlled study (Section~\ref{subsec:controlled_findings}). Several participants contrasted ParseJargon favorably with their usual strategies for handling unfamiliar terms, such as waiting for more context or searching the web during the meeting. For example, one participant noted that \textit{"It saved me the hassle of trying to Google stuff as the meeting went on"} (P3), while another said \textit{"I like that it keeps me engaged with the speaker... it made me want to put more effort into understanding jargon in presentations"} (P7).

\subsubsection{Practical Limitations}
At the same time, the deployment surfaced two practical limitations. First, two participants reported low willingness for future use and low usefulness in comprehension because of inaccuracies in jargon identification. One of these two participants acknowledged the design of the system but shared concerns of identification correctness, \textit{"The correct level of identification is not good. If it worked properly, I think it would be helpful."} Most of these inaccuracies appeared to come from caption errors generated by the platform's speech-to-text service, potentially exacerbated by environmental noise or accent differences. Second, this was a within-team meeting in which some participants already knew much of the terminology, reducing the need for support in the first place. As one participant noted, \textit{"It was not very useful for me and provided no real new information. The level of jargon in the meeting was not too high for me because I already knew most of the content"}. Together, these observations suggest that reliable captions and sufficient cross-domain meeting content are important conditions for perceived value and trust of our system in real-world use.
\section{Discussion}
\label{sec:discussion}


With the growing integration of AI into knowledge work, it is important to determine when and where automated support can smooth communication barriers that traditionally require substantial manual effort to overcome. Prior work on unfamiliar-term support has primarily focused on written or recorded content \citep{semantic_reader, 10.1145/3613904.3642573, Bao2025, liu2025exploringdesignspacerealtime}. Our work extends this line of research by showing that personalization can make real-time jargon support more useful in meetings by reducing the information overload caused by generic assistance, and that practical personalization can be achieved through lightweight techniques, instead of heavy user modeling. Below we discuss promising directions that further enhance personalized support and extend beyond live meeting usage.


\subsection{Enhancing Personalized Jargon Support} \label{subsec:enhancing_personalization}

Our current system provides personalization on \emph{which} terms are surfaced. A next step is to personalize the explanations themselves \citep{Murthy2021TowardsPD}. Different listeners may want different levels of detail, technicality, or supporting examples, even when they agree that a term is worth explaining. Future systems could use the same lightweight feedback and portable profile state introduced in ParseJargon to gradually adapt explanation depth or style, or give users more direct control through mechanisms such as a complexity slider for adjusting explanation detail~\cite{personalization_slider}. For instance, domain experts could receive concise, technical definitions, whereas users less familiar with the topic might benefit from short examples or analogies tailored to their backgrounds~\cite{tal_audience_2024}. Prior work on prerequisite terms \citep{li2019should} also suggests an interesting middle ground: rather than surfacing chains of related concepts during the meeting, systems could provide deeper scaffolding in a post-meeting summary.

Explanation format is another direction. ParseJargon currently relies on concise text, which works well for low-disruption support, but some concepts may be better explained with figures, tables, or small interactive visualizations~\cite{CrossTalk2023, liu2025exploringdesignspacerealtime}. To balance utility and attention, explanations could follow a progressive path: brief, non-intrusive inline hints in the moment, with expandable details pinned to the sidebar glossary for later review. Given current research on generative user interfaces \citep{10.1145/3706598.3713285}, one exciting possibility is not only generating textual definitions in real time, but also providing multi-modal augmentations to the definition.

In addition, pilot deployment feedback highlighted the importance of organization-specific jargon. Future systems may therefore need organization-aware resources, such as company-maintained glossaries or retrieval over internal documentation, to improve precision and reduce hallucinations \citep{asai2024openscholar}. At the same time, while ambient explanations can lower the barrier to understanding, they should not replace human clarification entirely. Future systems might encourage follow-up questions to the speaker, or surface likely sources of confusion to speakers themselves, so that jargon support complements rather than suppresses open discussion, especially in workplaces.

Finally, the refined ParseJargon also takes a first step toward longer-term support through portable cross-session profile reuse. In our simulation, in-session feedback produced a larger improvement than cross-session reuse, while combining both yielded only a marginal gain beyond feedback alone. This suggests that immediate corrective control is the more impactful refinement at the current scale, whereas reusable memory becomes more valuable over longer time spans and richer histories. Over time, such a system could evolve toward a longer-term personal knowledge assistant that separates explicit self-description from accumulated evidence of what a user already knows, what kinds of explanations they find helpful, and which terms recur across contexts~\cite{adaptive_hyper, 10.1145/3544548.3580900, 10.1145/3290605.3300233}. The main challenge is to make such memory both useful and inspectable, so users remain in control of what is remembered and reused.

\subsection{Extending Beyond Live Meetings}
\label{subsec:broader_contexts}

Although ParseJargon was designed for live meetings, the same approach may be valuable in asynchronous settings. Because our controlled study relied on pre-recorded presentations, it already suggests a natural extension to videos, aligning with recent efforts that explore jargon assistance in video content~\cite{liu2025exploringdesignspacerealtime}. Specifically, ParseJargon could be effectively integrated into educational environments such as recorded lectures, webinars, or training sessions. Learners typically engage independently with specialized content and lack immediate opportunities to seek clarification. Integrating ParseJargon into recorded materials could provide learners with on-demand personalized jargon explanations, especially when learners have diverse knowledge background, allowing them to remain focused on core content without repeated context switching interruptions to search for definitions.

Another compelling extension is to support presenters rather than listeners. 
Presenters frequently struggle to balance technical depth with accessibility, especially in interdisciplinary or public-facing scenarios \citep{tullis2023curse, swords2023science}. ParseJargon could enable speakers to anticipate which terms may require preemptive clarification based on hypothetical or known audience profiles. By identifying terms that likely lead to comprehension challenges, speakers could proactively refine their content for clarity and accessibility before the speech. Such functionality would be valuable for both live presentations and recorded content, enhancing communication effectiveness and engagement across varied audiences.

\section{Limitations and Future Work}
\label{sec:limitations}

While the controlled study, system evaluation, and pilot deployment provide complementary evidence, several limitations remain. First, all empirical studies involved relatively small samples from a single technology company, which may limit generalizability to other organizational cultures and expertise distributions. Future work should test ParseJargon in larger and more diverse settings, especially in meetings with stronger cross-domain knowledge gaps. Second, our system evaluation is simulation-based. We simulated in-session feedback and cross-session reuse using post-hoc \textit{helpful} \& \textit{not helpful} labels from the controlled study. This offers a practical estimate of how the refined personalization mechanisms may behave, but it cannot fully capture how people would provide feedback in the moment, how their profiles evolve over time, or how portable memory accumulates across many meetings. Future work should therefore examine longitudinal real-world use. Finally, the system relies on external speech-to-text and LLM services. Variability in caption accuracy and timing can propagate to defining jargon in real time. Our architecture also processes each caption chunk individually, and we did not evaluate gating strategies (e.g., user-triggered or signal-based invocation), batching, or caching that could reduce computational cost and latency. Exploring speech recognition robustness, gated/cached pipelines are important next steps. Improving robustness to transcription errors and reducing latency remains an important direction for future work. Looking ahead, newer language models and speech-to-text services with higher accuracy and lower latency may directly help mitigate some of these limitations.
\section{Conclusion}
\label{sec:conclusion}

A persistent stumbling block for effective communication is knowing what a listener knows \citep{tullis2023curse}. This ``curse of knowledge'' is often revealed through specialized vocabulary that speakers use fluently but listeners may not understand. In this paper, we take a first step toward addressing this challenge in real time with ParseJargon, a system for personalized jargon support in online meetings. We began with an initial prototype that used a simple one-sentence profile and a two-stage identify--then--filter pipeline. A controlled study showed that even this minimal personalization improves comprehension and perceived value of others' work while avoiding information overload to maintain engagement. Guided by these findings, we then refined ParseJargon with a richer profile interface, in-session per-term feedback, and a persistent glossary with portable cross-session reuse. Simulation-based evaluation showed that these refinements further improved jargon-selection precision, while latency analysis and live deployment suggested that the system remains feasible and usable for real-time meetings. Taken together, these results position ParseJargon as a practical path toward personalized language support that helps diverse audiences follow---and appreciate---work across domains.


\bibliographystyle{ACM-Reference-Format}
\bibliography{ref}

\appendix
\appendix

\twocolumn

\section{Prompts and Parameters}
\label{app:prompt}

Parameters: temperature is set to 0.1 and maximum length is 1000.

\subsection{Jargon Identification \& Explanation}

\begin{description}
   \item[System Message] \small\texttt{Your job is to help a listener understand speeches that might contain jargon terms they are unfamiliar with. You will be given the transcript snippet. For each snippet, the format will be "Transcript: [snippet]". Your task is to first identify any of those terms that the listener might not fully understand, then provide a definition for each term in concise plain language. Your output should be in the format of a list of term-definition pairs. Return only valid JSON in the format [\{"term": "definition"\}, ...]. Do not include additional commentary or text outside the JSON. Leave the list blank if you think all the terms in the input transcript are common words that do not need additional explanations. Do not include terms that are already in the previously defined term list.}\normalsize

   \item[User Prompt] \small\texttt{Transcript: \{\textit{transcript}\}, Previously defined terms: \{\textit{defined\_terms}\}}\normalsize
\end{description}

\subsection{Personalization}

\begin{description}
   \item[System Message] \small\texttt{You are given a glossary, a user profile, and a user preference list. The user preference list may include terms previously marked as Helpful or Already Knew. Your job is to remove terms the user is likely to already understand based on both their profile and preference list. Terms marked as Already Knew are strong evidence that the user likely already understands that term and therefore may prefer fewer similar terms. Terms marked as Helpful indicate that support on that term was useful and should not be used as evidence to suppress related terms. The input glossary is provided in valid JSON format, where each item is structured as \{"term": "definition"\}. Examine only the terms (the keys in the JSON) and remove the terms the user is likely already familiar with from the glossary. Return only valid JSON structured exactly as: \{"understood\_terms": ["term1", "term2", ...], \newline"refined\_glossary": [\{"term": "definition"\}, ...]\}. Do not include any extra commentary or text.}\normalsize

   \item[User Prompt] \small\texttt{Glossary: \{\textit{glossary}\}, User Profile: \{\textit{profile}\}, User preference: \{\textit{preferences}\}}\normalsize
\end{description}

\paragraph{Implementation Note.}
In the current implementation, ParseJargon uses the same personalization prompt for both in-session feedback and cross-session reusable history. Imported profile and glossary history can be appended to \textit{profile} and \textit{preferences} using the same term-level format.

\newpage

\section{Controlled Study}

This section includes four tables to show 1) the demographic information for controlled study participants, 2) the presentation viewing order for the controlled study, 3) test statistics for all metrics in the controlled study, and 4) example takeaways and questions in the controlled study.

\begin{table}[h]
\small
  \caption{Controlled experiment participant profiles, with participant indices hidden and rows reordered for anonymity}
  \begin{tabular}{ll}
    \toprule
    \textbf{Education}&\textbf{Job Role} \\
    \midrule
    Statistics PhD & Machine Learning Researcher \\
    Computer Science Master & Research Engineer \\ 
    Applied Mathematics PhD & Oceanography Researcher \\
    Computer Science Master & Data Engineer \\
    Physics PhD & Quantum Researcher \\
    Civil Engineering PhD & Earth Science Researcher \\
    Computer Science Bachelor & Application Engineer \\
    \bottomrule
  \end{tabular}
  \label{tab:participants}
\end{table}


\begin{table}[h]
\small
  \caption{Viewing schedule for every participant, using a counter-balanced design.
  Entry format = \textit{condition–presenter}.  
  Conditions: 0 Baseline, 1 Generic, 2 Personalized.}
  \label{tab:counterbalance}
  \centering
  \begin{tabular}{c|ccc|ccc}
    \toprule
    \multirow{1}{*}{Audience} & \multicolumn{3}{c|}{Session 1} & \multicolumn{3}{c}{Session 2}\\
    \midrule
    P1 & 0–P2 & 1–P3 & 2–P4 & 1–P5 & 2–P6 & 0–P7\\
    P2 & 1–P3 & 2–P4 & 0–P5 & 2–P6 & 0–P7 & 1–P1\\
    P3 & 2–P4 & 0–P5 & 1–P6 & 0–P7 & 1–P1 & 2–P2\\
    P4 & 0–P5 & 1–P6 & 2–P7 & 1–P1 & 2–P2 & 0–P3\\
    P5 & 1–P6 & 2–P7 & 0–P1 & 2–P2 & 0–P3 & 1–P4\\
    P6 & 2–P7 & 0–P1 & 1–P2 & 0–P3 & 1–P4 & 2–P5\\
    P7 & 0–P1 & 1–P2 & 2–P3 & 1–P4 & 2–P5 & 0–P6\\
    \bottomrule
  \end{tabular}
\end{table}

\begin{table*}[h]
\small
  \caption{Test statistics for all metrics, including $w$ (Wilcoxon signed-rank test), corrected $p$-value for $w$, and  Cohen’s $d$. Holm–Bonferroni method was used for post-hoc correction.}
  \label{tab:full-stats-selfreport}
  \centering
  \setlength\tabcolsep{5pt}
  \begin{tabular}{l|lll|lll|lll}
    \toprule
      & \multicolumn{3}{c}{Generic vs Baseline}
      & \multicolumn{3}{|c}{Personalized vs Baseline}
      & \multicolumn{3}{|c}{Personalized vs Generic}\\
      \cmidrule(lr){2-4} \cmidrule(lr){5-7} \cmidrule(lr){8-10}
      Metric &
      $w$ & $p_w$ & $d$ &
      $w$ & $p_w$ & $d$ &
      $w$ & $p_w$ & $d$ \\
    
    \midrule
    Comprehension &
    56.0 & 0.0294 & 0.6682  &
    86.5 & 0.0047 & 1.2455  &
    24.5 & 0.0294 & 0.5316 \\

    Engagement &
    26.0 & 0.7395 & -0.1988  &
    31.5 & 0.2733 & 0.2938  &
    25.5 & 0.0710 & 0.5942 \\

    Value &
    48.0 & 0.0658 & 0.4242  &
    72.5 & 0.0073 & 1.1127  &
    21.0 & 0.0196 & 0.7687 \\

    Usefulness$^\dag$ &
    -- & -- & --  &
    -- & -- & --  &
    50.5 & 0.0065 & 0.8654 \\

    \midrule
    Clarity &
    49.0 & 0.3215 & 0.0626  &
    59.0 & 0.2248 & 0.2673  &
    59.0 & 0.3232 & 0.1797 \\

    Relevance &
    90.0 & 0.0076 & 0.7751  &
    91.0 & 0.0055 & 0.8832  &
    40.0 & 0.5527 & -0.0626 \\

    Depth &
    68.0 & 0.1551 & 0.3378  &
    86.0 & 0.0099 & 0.7751  &
    65.0 & 0.0647 & 0.4493 \\
    \bottomrule
  \end{tabular}

  \vspace{3pt}
  {\raggedright\footnotesize
  $^\dag$ Usefulness is only defined for Generic and Personalized conditions. \par}
   \vspace{3em}
\end{table*}

\renewcommand{\arraystretch}{1.1}
\begin{table*} [h]
\small
  \caption{Example takeaways and questions from different glossary conditions, illustrating differences in relevance and depth (presentation topic: deep learning in earth science (1) \& predicting ocean surface properties (2))}

  \begin{tabular}{  m{7em} | m{35em} } 
    \toprule
    \textbf{Condition}&\textbf{Example Takeaways} \\
    \midrule
    Baseline & \textit{(1) Foundation models help drastically improve accuracy of geospatial detection.}\\
     & \textit{(2) Using AI to understand behaviour of ocean surfaces.}\\
    \hline
    Generic & \textit{(1) We are using AI for climate challenges and modeling.}\\
     & \textit{(2) Using neural networks with Fourier Transforms to predict ocean depths is pretty reliable.}\\
    \hline
    Personalized & \textit{(1) Finetune some large pretrained neural nets for geospatial images classification and segmentation.}\\
     & \textit{(2) FNO is versatile in terms of accepting the input and output of different sizes to predict the ocean surface properties.}\\
    \midrule
    \midrule
    \textbf{Condition}&\textbf{Example Questions} \\
    \midrule
    Baseline & \textit{(1) How did the model deal with the different kinds of data sets?}\\
     & \textit{(2) What are the implications of your study to real life?}\\
    \hline
    Generic & \textit{(1)What are the key importance of using these models together, versus those designed for specific problems independently if they have already been trained?}\\
     & \textit{(2) How does extending the prediction accuracy time help the environmental scientist or decision makers in general?}\\
    \hline
    Personalized & \textit{(1) Isn’t segmentation fundamentally also classification task? How different is the method used for segmentation vs classification?}\\
     & \textit{(2) Could you elaborate more on spectrum space and what’s the benefit of using spectrum loss function over (traditional) cross entropy loss?}\\
    \bottomrule
  \end{tabular}
  \label{tab:comprehension_example}
\end{table*}


\end{document}